\documentclass[conference]{IEEEtran}
\IEEEoverridecommandlockouts
\usepackage{cite}
\usepackage{amsmath,amssymb,amsfonts}
\usepackage{algorithmic}
\usepackage{graphicx}
\usepackage{textcomp}
\usepackage{xcolor}
\usepackage{hyperref}
\usepackage{booktabs}
\usepackage{makecell}

\def\BibTeX{{\rm B\kern-.05em{\sc i\kern-.025em b}\kern-.08em
    T\kern-.1667em\lower.7ex\hbox{E}\kern-.125emX}}

\begin{document}

% ---------------------------------------------------------------
% 1. Title
% ---------------------------------------------------------------
\title{CogSR: Semantic-Aware Speech Super-Resolution via Chain-of-Thought Guided Flow Matching}

% Double-blind review: Author information must be hidden for submission.
% Uncomment the following block for the camera-ready version.
 \author{
     \IEEEauthorblockN{Jiajun Yuan\IEEEauthorrefmark{1}, Xiaochen Wang\IEEEauthorrefmark{1}, Yuhang Xiao\IEEEauthorrefmark{1}, Yulin Wu\IEEEauthorrefmark{2}, Chenhao Hu\IEEEauthorrefmark{3} and Xueyang Lv\IEEEauthorrefmark{3}}
     \IEEEauthorblockA{\IEEEauthorrefmark{1}National Engineering Research Center for Multimedia Software, School of Computer Science, Wuhan University, Wuhan, China}
     \IEEEauthorblockA{\IEEEauthorrefmark{2}School of Artificial Intelligence, Jianghan University, Wuhan, China}
    \IEEEauthorblockA{\IEEEauthorrefmark{3}Xiaomi Corporation, Beijing, China}
    \IEEEauthorblockA{E-mails: yjj2002@whu.edu.cn}
 }
%\author{\IEEEauthorblockN{Anonymous ICME Submission}}

\maketitle

% ---------------------------------------------------------------
% 2. Abstract
% ---------------------------------------------------------------
\begin{abstract}
Applying speech super-resolution (SR) to recordings with severely low sampling rates is a critical challenge in digital archiving and investigative audio recovery. In these scenarios, the input lacks essential acoustic cues. Consequently, existing generative models often fail; without sufficient context, they hallucinate phonetic content, guessing words based on probability rather than meaning.
 To address this, we propose CogSR, a framework designed specifically for high-precision, offline restoration. Our approach shifts the focus from simple signal mapping to cognitive reconstruction. By integrating a Large Audio-Language Model, we employ Chain-of-Thought reasoning to act as a semantic anchor, while explicit acoustic priors ensure the speaker's identity remains consistent. This guides a Rectified Flow backbone to synthesize high-frequency details that are not only realistic but linguistically accurate. Evaluations show that CogSR effectively eliminates ambiguity in severe degradation regimes, making it a robust solution for restoring high-value legacy and surveillance audio.
\end{abstract}

% ---------------------------------------------------------------
% 3. Keywords
% ---------------------------------------------------------------

% ---------------------------------------------------------------
% 4. Introduction
% ---------------------------------------------------------------
\section{Introduction}

\label{sec:intro}
Speech super-resolution (SR) aims to recover high-fidelity audio from low-resolution recordings, thereby bridging the gap between legacy media and modern listening standards. By doing so, it plays an indispensable role in a wide range of applications, including digital archiving, digital media restoration, and audio post-production. In particular, restoring historical archives and legacy media is of paramount importance; these recordings often hold irreplaceable cultural and personal value but are often plagued by severe quality degradation \cite{liu2023audiosr,liu2022voicefixer}.

\begin{figure}[t]
    \centering
    \includegraphics[width=0.8\columnwidth]{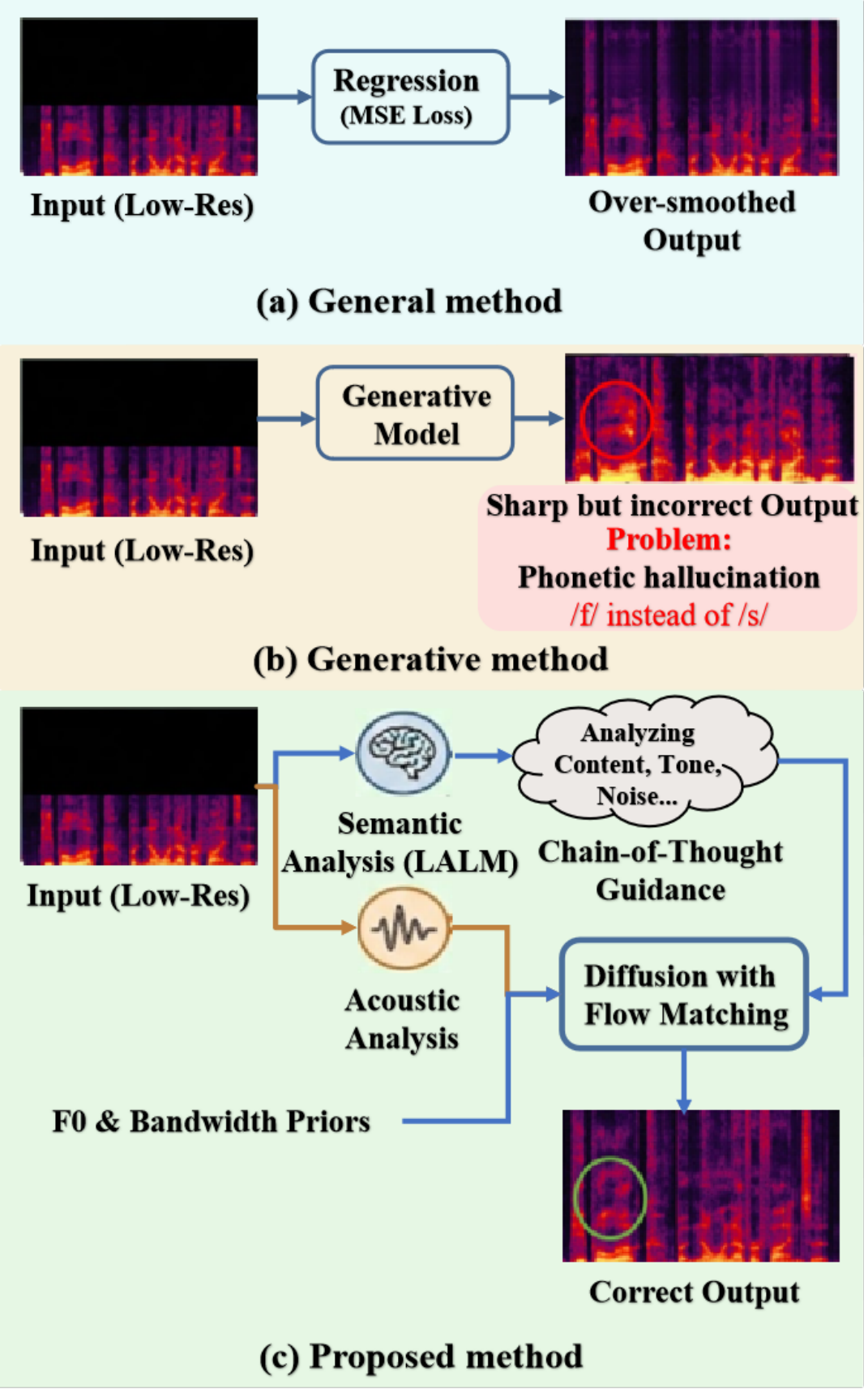} 
    \vspace{-0.3cm}
    \caption{Visual comparison of different speech restoration paradigms under severe degradation.}
    \vspace{-0.3cm}
    \label{fig:teaser}
\end{figure}

Historically, speech super-resolution was predominantly treated as a deterministic regression task, where deep neural networks were trained to minimize reconstruction errors such as Mean Squared Error (MSE). Though these methods excel at recovering the spectral envelope, they inevitably produce over-smoothed, muffled textures—and crucially, fail to reconstruct the fine-grained high-frequency details that are essential for perceptual realism \cite{wang2015speech}. To address this shortcoming, the field has since shifted toward a new paradigm: generative modeling. Methods leveraging Generative Adversarial Networks (GANs), and more recently those based on Diffusion Probabilistic Models (DPMs), have shown great promise here. Unlike point-wise estimation, these generative frameworks model the data distribution directly, enabling them to synthesize rich, realistic spectral details that significantly elevate the perceptual quality of restored speech \cite{evans2024stable}.

However, a fundamental limitation arises when these generative systems are applied 
to low sampling rate scenarios. 
In such regimes, critical cues required for phoneme distinction are entirely lost. 
Existing models, which typically rely on low-level features or simple text prompts, 
lack the capacity to infer missing semantic content accurately. 
Consequently, they act as blind guessers, 
hallucinating plausible-sounding but semantically incorrect high-frequency content based on statistical priors. 
As illustrated in Fig. \ref{fig:teaser}, 
this leads to phonetic hallucinations and prosodic inconsistencies, 
which are unacceptable for content-sensitive applications.

To address this challenge, we propose that robust speech restoration must shift from traditional signal mapping to cognitive reconstruction—meaning the restoration process must be centered on a deep semantic understanding of the spoken content. Guided by recent advances in Large Audio-Language Models, we further introduce a mechanism where the model explicitly reasons about the underlying attributes of the speech signal before generating output, and this thought directly informs the design of the framework proposed in this paper.

Building on this, we present CogSR, a semantically aware speech super-resolution framework via latent flow matching, whose core innovation lies in its Chain-of-Thought (CoT) guidance mechanism. Unlike conventional static captions, we employ Qwen2-Audio \cite{chu2024qwen2} as a cognitive agent, which decomposes the input into three structured dimensions (linguistic content, paralinguistic emotion, and environmental context) following a structured cognitive protocol. These analytical results are then encoded by T5-base to serve as robust semantic anchors, effectively bridging the gap between high-level semantic understanding and low-level signal generation, and fundamentally preventing the model from hallucinating incorrect phonemes. Meanwhile, to ensure the synthesized high frequencies align with the speaker’s original voice, we additionally incorporate parametric fundamental frequency and bandwidth constraints.

Technically, CogSR is built on a Diffusion Transformer backbone, trained using a Rectified Flow objective \cite{lipman2023flow}. This formulation enables the construction of straight-line probability paths, significantly enhancing training stability and generation quality. By deeply integrating cognitive reasoning with generative flow matching, our system ultimately achieves synergistic optimization of semantic accuracy and signal fidelity, effectively resolving the core challenges previously faced by speech super-resolution methods.

The main contributions of this paper are summarized as follows:
\begin{itemize}
    \item \textbf{Chain-of-Thought Semantic Guidance:} We propose a novel conditioning mechanism 
    where a Large Audio-Language Model acts as a reasoning agent 
    to analyze deep semantics and speech properties. 
    This structured analysis acts as a semantic anchor, 
    effectively mitigating linguistic hallucinations in low sampling rate scenarios.
    \item \textbf{Speech-Specific Priors:} We introduce an explicit guidance module 
    that conditions the generative process on fundamental frequency ($F_0$) and bandwidth priors. 
    This design enforces constraints specific to human vocal production, 
    ensuring the preservation of speaker identity and natural prosody, 
    which are often neglected by general audio models.
    \item \textbf{Unified Latent Flow Framework:} We present a cohesive system architecture 
    that integrates cognitive reasoning with Rectified Flow Matching in a compressed latent space. 
    Extensive evaluations demonstrate that this framework achieves state-of-the-art performance 
    in terms of speech intelligibility and perceptual quality compared to existing baselines.

\end{itemize}

% ---------------------------------------------------------------
% 5. Related Work
% ---------------------------------------------------------------
\section{Related Work}
\label{sec:related}

\subsection{Speech Super-Resolution and Restoration}
Speech super-resolution (SR) has evolved from signal processing-based estimation to deep learning-driven restoration. Early neural-based methods, for instance, framed SR as a regression task: they optimized $L_1$ or MSE losses to map low-resolution (LR) spectral features to their high-resolution (HR) counterparts \cite{li2015dnn, wang2015speech}. While these discriminative models excel at recovering the spectral envelope, they tend to produce overly smoothed speech textures—ultimately leading to muffled perceptual quality \cite{liu2022voicefixer}.

To boost the realism of synthesized speech, researchers turned to generative models, starting with Generative Adversarial Networks (GANs). Approaches like mdctGAN \cite{shuai23_interspeech} and HiFISR \cite{zhao2025hifisr}, for example, use adversarial training to reconstruct fine-grained spectral details. Separately, another line of work focused on neural vocoder-based methods: models such as NVSR \cite{liu2023neural} and VoiceFixer \cite{liu2022voicefixer} leverage pre-trained vocoders to resynthesize high-fidelity waveforms from predicted mel-spectrograms. Yet for all their improvements in signal fidelity, these generative models share a critical limitation: In low sampling rate scenarios where key phonemic cues are missing, they lack the high-level semantic guidance needed to resolve ambiguous phonemes, often resulting in linguistic hallucinations and incorrect lexical content \cite{atwany2025hallucination}.
\subsection{Flow Matching for Generative Speech Modeling}
Diffusion Probabilistic Models (DPMs) have shown remarkable capabilities in speech synthesis and enhancement, but a key barrier to their practical deployment is the high computational cost of iterative sampling. Flow Matching (FM) \cite{lipman2023flow} and Rectified Flow \cite{liu2022flow} has emerged as a more efficient alternative, which models deterministic straight-line trajectories between noise and data distributions. This paradigm enables high-fidelity speech generation with far fewer sampling steps.

In the speech domain, Flow Matching has been successfully applied to text-to-speech synthesis, including VoiceFlow \cite{guo2024voiceflow}, Matcha-TTS \cite{mehta2024matcha} and speech enhancement \cite{wang2025flowse}. These works demonstrate FM’s ability to effectively capture the complex conditional distributions of speech signals, such as prosody and timbre. However, applying Flow Matching to restoration tasks especially under severe degradation, where semantic guidance remains an under-explored area. Unlike TTS models that generate speech from scratch, restoration models must strike a balance between fidelity to the original signal and hallucination-free reconstruction of missing high-frequency content: a challenge standard flow matching frameworks fail to explicitly address. Although CogSR leverages a Large Audio-Language Model (LALM) for Chain-of-Thought cues, this reasoning is executed offline and cached per utterance, so the online restoration path remains dominated by the lightweight 32-step Rectified Flow sampler. Consequently, our design preserves the sampling-efficiency advantages of flow matching while retaining the semantic fidelity delivered by LALM guidance.
\begin{figure*}[t]
    \centering
    \includegraphics[width=0.95\textwidth ]{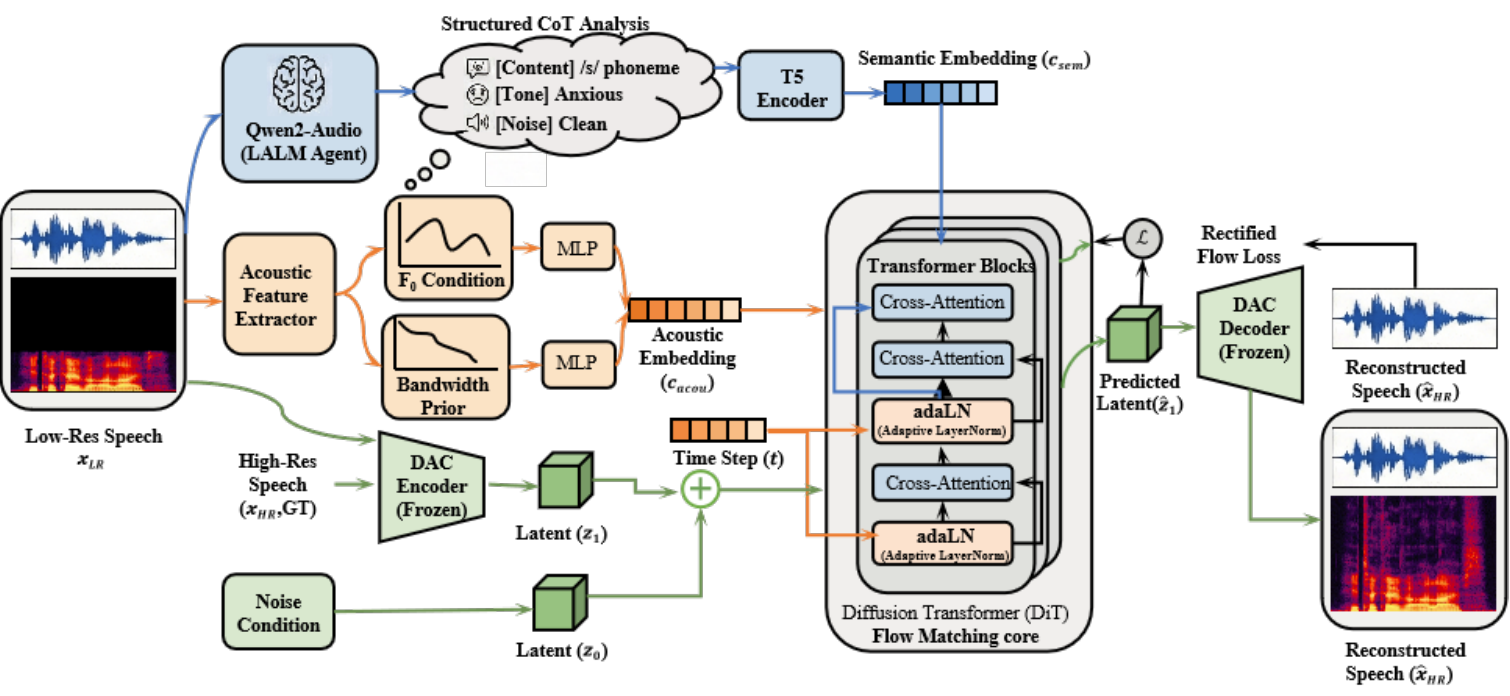}
    \vspace{-0.3cm}
    \caption{Overview of the proposed CogSR architecture.}
    \vspace{-0.3cm}
    \label{fig:arch}

\end{figure*}
\subsection{Large Audio-Language Models for Speech Reasoning}
The convergence of Large Language Models (LLMs) and audio processing has led to the development of Large Audio-Language Models (LALMs) capable of interpreting and reasoning about spoken content. Prominent architectures, such as Qwen-Audio \cite{chu2023qwenaudio} and Qwen2-Audio \cite{chu2024qwen2}, process audio inputs to address diverse comprehension tasks.

Recent research has further extended Chain-of-Thought (CoT) reasoning to the audio domain. Frameworks like CoT-ST \cite{du2024cotst} and Thinking-with-Sound \cite{xiong2025thinking} employ LALMs to generate intermediate reasoning steps, analyzing paralinguistic cues—including emotion, gender, and acoustic environment—prior to determining the final output. In these settings, the LALM functions as a cognitive agent that decomposes complex speech signals into structured semantic attributes. However, the use of such reasoning chains to guide low-level generative processes remains underexplored. While studies such as SCOUT \cite{li2025scout} utilize CoT primarily for comprehension or high-level reasoning, the potential of CoT to serve as a semantic anchor for mitigating hallucinations in signal restoration tasks has yet to be fully investigated.
% ---------------------------------------------------------------
% 6. Methodology
% ---------------------------------------------------------------
\section{Methodology}
\label{sec:methodology}

In this section, we present the proposed CogSR framework. 
The overall architecture, depicted in Fig. \ref{fig:arch}, 
operates within the compressed latent space of a pre-trained neural audio codec 
to reduce computational complexity while capturing high-level acoustic structures. 
CogSR is built upon a Diffusion Transformer (DiT) backbone trained with a Rectified Flow objective. 
To address the ill-posed nature of bandwidth extension, particularly in low sampling rate scenarios, 
we introduce a dual-stream guidance mechanism: 
(1) \textit{Cognitive Semantic Anchoring} via structured Chain-of-Thought (CoT) reasoning 
from a Large Audio-Language Model (LALM), and 
(2) \textit{Multi-Factor Acoustic Guidance} utilizing explicit bandwidth and pitch priors 
to enforce physical constraints.

\subsection{Preliminary: Latent Rectified Flow Matching}
\label{subsec:rfm}

Generative models for speech restoration aim to learn a mapping from a simple prior distribution to the complex data distribution of high-fidelity speech. While Denoising Diffusion Probabilistic Models (DDPMs) have achieved promising results, they are often bottlenecked by the need for expensive iterative sampling. Flow Matching (FM) \cite{lipman2023flow} overcomes this limitation with a more general and efficient framework, regressing a velocity field $v_t(x)$ that transports a probability density path from a source $p_0$ to a target $p_1$.

Specifically, we employ \textit{Rectified Flow} (RF) \cite{liu2022flow}, a streamlined instance of FM that enforces straight-line trajectories to facilitate fast and stable sampling. Our implementation operates within the latent space of a frozen Descript Audio Codec (DAC) \cite{kumar2024high}. By leveraging DAC's highly compressed yet perceptually lossless representation, we avoid the phase reconstruction challenges and high dimensionality inherent in traditional spectrogram-based approaches. This design shifts the modeling focus toward semantic and acoustic structure rather than low-level waveform details, thereby enhancing both training efficiency and generation fidelity.

Formally, let $x_1$ denote the high-resolution speech latent extracted by the DAC encoder, and $x_0 \sim \mathcal{N}(0, I)$ the source noise. We define a linear interpolation path $x_t = t x_1 + (1-t) x_0$, implying a constant target velocity field $u_t = x_1 - x_0$. Accordingly, the DiT backbone $v_\theta(x_t, t, c)$ is trained to predict this velocity by minimizing the mean squared error:
\begin{equation}
    \mathcal{L}_{RF}(\theta) = \mathbb{E}_{t, x_0, x_1} \left[ \| v_\theta(x_t, t, c) - (x_1 - x_0) \|^2 \right]
\end{equation}
where $t \sim \mathcal{U}(0,1)$. Such a formulation simplifies the optimization landscape, enabling high-quality generation with fewer steps using standard ODE solvers.

To ensure the model captures the specific restoration task without bias from unrelated domains, we train the DiT backbone from scratch. To preserve the robust representational capabilities of the pre-trained components, we freeze the DAC encoder/decoder, T5 text encoder, and Qwen2-Audio reasoning agent throughout training. Optimization is thus restricted to the DiT parameters and lightweight projection layers, effectively guiding the mapping from degraded latents to high-fidelity speech while anchoring on the frozen semantic and acoustic guidance.
\subsection{Cognitive Semantic Anchoring via CoT}
\label{subsec:cot}
A critical vulnerability of existing generative SR models in low sampling rate regimes is linguistic hallucination, the generation of intelligible but incorrect phonemes when acoustic cues are ambiguous. To mitigate this, we introduce \textit{Cognitive Semantic Anchoring}, leveraging the reasoning capabilities of LALMs to ensure semantic consistency.

We employ Qwen2-Audio \cite{chu2024qwen2} as a cognitive reasoner.It is worth noting that essential linguistic semantics are predominantly preserved in the low-frequency band (e.g., $<1$ kHz). Furthermore, Qwen2-Audio exhibits remarkable robustness to such bandwidth limitations, enabling accurate content inference even from severely degraded signals. 
Unlike SAGA-SR \cite{im2025sagasr}, which conditions on static captions, CogSR obtains proactive CoT rationales that justify the perceived timbre, affect, and noise context before generation. To obtain these richer cues, rather than generating a simple caption we instruct the model to perform a structured Chain-of-Thought (CoT) analysis that explicitly decomposes the input audio into distinct semantic dimensions:
\begin{enumerate}
    \item \textbf{Paralinguistic Inference:} Analyzing speaker attributes such as gender, emotion and environmental context like background noise type, crucial for reconstructing appropriate timbre and ambience.
    \item \textbf{Linguistic Content:} Transcribing the spoken content to provide strong phonetic constraints.
    \item \textbf{Perceived Quality Assessment:} Providing a high-level description of the overall audio clarity and perceived degradation types such as "muffled speech," "noisy background", acting as a global style indicator.
\end{enumerate}
To instantiate this, we employ a specific prompt template that forces the model to output these attributes sequentially. A typical CoT output $T_{CoT}$ fed into the text encoder looks like: 
\textit{``[Gender]: Male; [Emotion]: Anxious; [Noise]: Street traffic; [Content]: Help is on the way; [Quality]: Low bandwidth, muffled.''}
This structured format ensures that the semantic embedding $c_{sem}$ captures orthogonal acoustic dimensions explicitly.

The resulting structured description $T_{CoT}$ is encoded by a frozen T5 encoder into semantic embeddings $c_{sem}$. These embeddings are injected into the DiT via cross-attention layers, acting as semantic anchors that constrain the generative process to align with the linguistic and paralinguistic context of the input, effectively suppressing hallucinations.
% --- Placeholder for Spectrogram Figure ---
\begin{figure*}[t]
\centering
\includegraphics[width=0.95\textwidth]{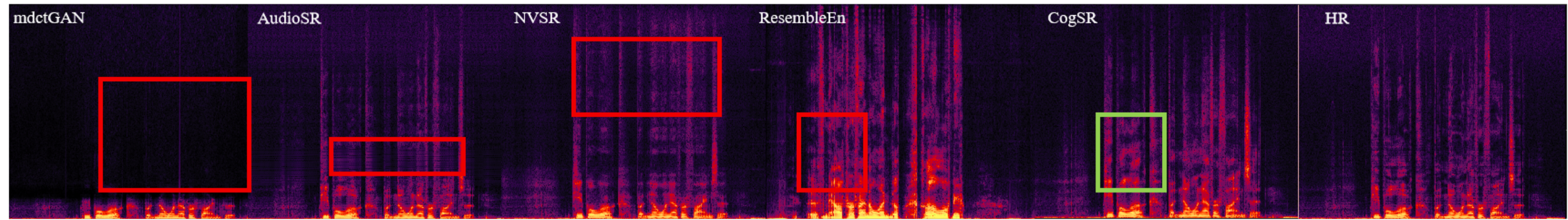}
\vspace{-0.3cm}
\caption{Comparison of restoration paradigms under severe degradation.}
\vspace{-0.3cm}
\label{fig:spectrograms}
\end{figure*}
\subsection{Multi-Factor Acoustic Guidance}
\label{subsec:acoustic}

Although semantic guidance maintains content consistency, it lacks the precision required for spectral reconstruction, particularly regarding speaker identity and prosody. To address this, we introduce a \textit{Multi-Factor Acoustic Guidance} module rooted in explicit physical priors.

We specifically model the effective signal bandwidth to define the cutoff frequency of the low-resolution input. By projecting this scalar into a high-dimensional embedding $c_{bw}$ via Fourier feature mapping, we provide a robust cue that prevents passband artifacts and preserves spectral continuity.

Furthermore, to mitigate metallic artifacts caused by harmonic misalignment, we incorporate prosodic features. Distributional $F_0$ characteristics are extracted via a robust pitch tracker (e.g., Parselmouth \cite{jadoul2018parselmouth}) and encoded into continuous embeddings $c_{pitch}$. These features impose structural constraints, aligning generated harmonics with the speaker's fundamental frequency.

These conditioning signals are integrated into the DiT architecture according to their distinct modalities. While the sequence-level semantic embedding $c_{sem}$ is injected via cross-attention, the acoustic priors ($c_{bw}$ and $c_{pitch}$) employ a dual-path mechanism. They are first projected into global embeddings and added to the time-step embedding to modulate the global trajectory; simultaneously, they are mapped to tokens and concatenated with $c_{sem}$ for cross-attention. This strategy enforces parametric adherence to bandwidth and identity constraints without suppressing local variations, ensuring the generated speech is both semantically accurate and acoustically authentic.

% ---------------------------------------------------------------
% 7. Experiments
% ---------------------------------------------------------------
\section{Experimental Setup}
\label{sec:experiments}

\subsection{Datasets and Data Preparation}
We trained CogSR on a combination of high-quality speech datasets: VCTK \cite{yamagishi2019cstr} and HiFi-TTS \cite{bakhturina2021hifi}. All audio data was resampled and unified at 44.1 kHz, totaling approximately 244 hours. The corpus was partitioned into training ($90\%$), validation ($5\%$), and testing ($5\%$) splits, ensuring speaker independence across subsets.

During training, we employed dynamic data augmentation to simulate real-world degradations. High-resolution (HR) audio was low-pass filtered with random cutoff frequencies sampled uniformly from $[2 \text{ kHz}, 16 \text{ kHz}]$ to simulate varying bandwidth limitations. Subsequently, to ensure robustness, the audio was corrupted by additive real-world environmental noise—rather than synthetic Gaussian noise—with signal-to-noise ratios (SNR) sampled randomly between $5 \text{ dB}$ and $15 \text{ dB}$.

For standardized evaluation, we defined a challenging benchmark featuring a fixed 4 kHz sampling rate combined with 5 dB SNR noise.

\subsection{Implementation Details}
CogSR leverages the compressed latent space of the pre-trained DAC (44.1kHz, 8kbps) \cite{kumar2024high}, with both the encoder and decoder kept frozen throughout experiments. During both training and inference, the Qwen2-Audio agent processes low-resolution (LR) inputs. This design compels the model to infer semantics directly from degraded signals, thereby bolstering robustness in real-world scenarios where high-resolution ground truth is unavailable.

Training used a global batch size of 128 on an NVIDIA A100. We optimized the RF objective using AdamW. For inference, we use the Euler ODE solver with 32 steps. 
\subsection{Baselines and Metrics}
We compare CogSR against a diverse set of state-of-the-art baselines representing different generative paradigms. For GAN-based approaches, we include mdctGAN \cite{shuai23_interspeech}, which operates on MDCT spectra, and NVSR \cite{liu2023neural}, which leverages a pre-trained neural vocoder for high-fidelity resynthesis. In the diffusion domain, we compare with AudioSR \cite{liu2023audiosr}, a latent diffusion model designed for universal audio super-resolution. We also include Resemble Enhance \cite{resemble2024enhance} as a representative flow matching-based speech enhancement model. Additionally, we report the performance of direct DAC Reconstruction \cite{kumar2024high} to serve as the theoretical upper bound for our latent-based framework.

Performance is assessed using a multi-dimensional metric suite focusing on semantic, spectral, and speaker fidelity. Word Error Rate (WER) is computed using the Whisper-large-v3 model \cite{radford2023robust} on noisy test utterances (5 dB SNR) to evaluate semantic intelligibility and robustness against hallucinations. Log-Spectral Distance (LSD) measures the physical spectral accuracy between the restored and ground truth signals, defined as:
\begin{equation}
    \text{LSD} = \frac{1}{T} \sum_{t=1}^{T} \sqrt{\frac{1}{F} \sum_{f=1}^{F} \left( \log_{10} \frac{S(t, f)^2}{\bar{S}(t, f)^2} \right)^2}
\end{equation}
where $S$ and $\bar{S}$ denote the amplitude spectra of the ground truth and generated speech, and $T, F$ represent time frames and frequency bins. Finally, Speaker Similarity (SIM) is calculated as the cosine similarity between embeddings extracted by WavLM-SV \cite{chen2022wavlm}, quantifying the preservation of the original speaker's timbre and prosody.

% ---------------------------------------------------------------
% 8. Results
% ---------------------------------------------------------------
\section{Results and Analysis}
\label{sec:results}

\subsection{Quantitative Comparison}
Table \ref{tab:objective_results} summarizes the quantitative performance on the VCTK test set across 4 kHz and 2 kHz sampling rates. Note that WER is assessed under noisy conditions (5 dB SNR) to gauge robustness, whereas LSD and SIM are calculated in clean environments to evaluate restoration fidelity.

\textbf{Word Error Rate.} Regarding intelligibility, CogSR achieves a WER of 4.20\% given 4 kHz inputs, surpassing baselines such as AudioSR (12.15\%) and NVSR (13.56\%) by a substantial margin while approaching the DAC topline (3.07\%). This robustness persists even in the extreme 2 kHz scenario: CogSR maintains a viable WER of 23.12\%, whereas the performance of baseline models collapses.

\textbf{Log-Spectral Distance.} In terms of spectral reconstruction, CogSR records an LSD of 0.91 (4 kHz). This result is competitive with leading systems and demonstrates significantly closer alignment to the topline (0.45) compared to alternative generative approaches.

\textbf{Speaker Similarity.} Driven by explicit $F_0$ and bandwidth priors, CogSR attains the highest speaker similarity (0.99) in the 4 kHz condition. This effectively matches the topline (1.00) and exceeds standard denoising or enhancement baselines, underscoring the capacity of structured acoustic guidance to preserve original timbre and prosody where unconstrained methods fail.

In our experiments, we observed that our system generates approximately 1 second of audio per 1.0 second of processing time on a single NVIDIA A40, making it highly cost-effective for quality restoration. The total memory footprint during inference is approximately 24GB, with the frozen LALM occupying 20GB.

% --- Table 1: Main Results (4 kHz and 2 kHz input sampling rates) ---
\begin{table}[t]
    \centering
    \caption{Objective evaluation on VCTK test set at 2 {kHz} and 4 {kHz} input sampling rates. WER is measured under 5 {dB} SNR noise.}
    \label{tab:objective_results}
    \resizebox{\columnwidth}{!}{
    \begin{tabular}{l|ccc|ccc|c}
        \toprule
        & \multicolumn{3}{c|}{\textbf{2 kHz input sampling rate}} & \multicolumn{3}{c|}{\textbf{4 kHz input sampling rate}} & \textbf{Parameters} \\
        \cmidrule(r){2-4} \cmidrule(l){5-7}
        \textbf{Method} & \textbf{WER} ($\downarrow$) & \textbf{LSD} ($\downarrow$) & \textbf{SIM} ($\uparrow$) & \textbf{WER} ($\downarrow$) & \textbf{LSD} ($\downarrow$) & \textbf{SIM} ($\uparrow$) & \\
        \midrule
        % [WER, LSD, SIM] @ 4 kHz  |  [WER, LSD, SIM] @ 2 kHz
        DAC Rec& 3.07\% & 0.63 & 0.88 & 3.07\% & 0.45 & 1.00 & -\\
        Qwen2-Audio \cite{chu2024qwen2} & 1.00\% & - & - & 1.00\% & - & - & 7B\\
        \midrule
        mdctGAN \cite{shuai23_interspeech} & 43.34\% & 1.70 & 0.47 & 37.5\% & 0.92 & 0.74 & 101.0M\\
        NVSR \cite{liu2023neural} & 59.53\% & 1.04 & 0.68 & 13.56\% & 0.91 & 0.80 & 99.0M\\
        AudioSR(Speech) \cite{liu2023audiosr} & 35.69\% & 1.05 & 0.61 & 12.15\% & 1.26 & 0.82 & ~400M\\
        Resemble Enhance \cite{resemble2024enhance} & 31.91\% & 1.21 & 0.74 & 4.85\% & 0.95 & 0.97 & -\\
        \midrule
        \textbf{CogSR (Ours)} & \textbf{23.12\%} & \textbf{1.01} & \textbf{0.87} & \textbf{4.20\%} & \textbf{0.91} & \textbf{0.99} & \makecell{7.3B\textsuperscript{*}\\(145M Trainable)}\\
        \bottomrule
    \end{tabular}
    }
    \textsuperscript{*}Most parameters are from the frozen Qwen2-Audio and T5 models.
    \vspace{-0.2cm}
\end{table}

\subsection{Subjective Evaluation}

To assess perceptual fidelity in the 4 kHz-to-44.1 kHz bandwidth extension task, we conducted a Mean Opinion Score (MOS) study evaluating Overall Quality (MOS-Q) and Intelligibility (MOS-I) on a 5-point scale. As detailed in Table \ref{tab:subjective_results}, CogSR outperforms all competing methods. Significantly, it achieves parity with the hidden Ground Truth in intelligibility (4.60) and closely rivals the reference in overall quality (4.20 vs. 4.25). These results validate the critical role of semantic guidance in synthesizing speech that is both accurate and perceptually natural.

\begin{table}[h]
    \centering
    \vspace{-0.3cm}
    \caption{MOS test results for speech super-resolution (4 kHz $\to$ 44.1 kHz).}
    \label{tab:subjective_results}
    \resizebox{0.95\columnwidth}{!}{
    \begin{tabular}{lcc}
        \toprule
        \textbf{Method} & \textbf{MOS-Q} (CI) $\uparrow$ & \textbf{MOS-I} (CI) $\uparrow$ \\
        \midrule
        AudioSR \cite{liu2023audiosr} & $3.75 \pm 0.06$ & $4.30 \pm 0.06$ \\
        NVSR \cite{liu2023neural} & $4.00 \pm 0.06$ & $4.50 \pm 0.06$ \\
        Resemble Enhance \cite{resemble2024enhance} & $4.05 \pm 0.07$ & $4.50 \pm 0.07$ \\
        \midrule
        \textbf{CogSR (Ours)} & $\textbf{4.20} \pm \textbf{0.06}$ & $\textbf{4.60} \pm \textbf{0.06}$ \\
        \midrule
        Ground Truth  & $4.25 \pm 0.06$ & $4.60 \pm 0.06$ \\
        \bottomrule
    \end{tabular}
    }
    \vspace{-0.2cm}
\end{table}

\subsection{Qualitative Analysis}
We provide a visual comparison of spectrograms in Fig. \ref{fig:spectrograms}. 
The 4 kHz sampling rate input, resampled to 44.1 kHz, lacks all high-frequency content. 
The GAN-based mdctGAN produces an over-smoothed spectrum, missing fine-grained textures. 
The blind generative model Resemble Enhance hallucinates incorrect formants (highlighted in red box), 
changing the phoneme. 
In contrast, our CogSR, guided by semantic CoT, 
accurately reconstructs the high-frequency fricative structure (green box) consistent with the ground truth, 
while maintaining a natural harmonic structure thanks to acoustic priors.

% ------------------------------------------

\subsection{Ablation Study}
We conducted an ablation study on the 4 kHz sampling rate test set (Table \ref{tab:ablation}).

\begin{enumerate}
    \item \textbf{w/o CoT:} Replacing the structured CoT with ordinary text or complete omission. This confirms that explicit, structured semantic reasoning is crucial for accurate phoneme restoration when acoustic cues are ambiguous.
    \item \textbf{w/o Acoustic Priors:} Removing explicit $F_0$ and bandwidth embeddings 
    results in a strong degradation in Speaker Similarity
    and slightly worse spectral fidelity . 
    This indicates that while semantic guidance fixes content, 
    acoustic priors are essential for preserving speaker identity and stable harmonics.
\end{enumerate}

% --- Table 2: Ablation Study ---
\begin{table}[h]
\vspace{-0.3cm}
\caption{Ablation study for 4 kHz sampling rate speech SR.}
\label{tab:ablation}
\centering
\begin{tabular}{l|c|c|c}
\toprule
\textbf{Variant} & \textbf{WER} $\downarrow$ & \textbf{LSD} $\downarrow$ & \textbf{SIM} $\uparrow$ \\
\midrule
\textbf{CogSR (Full)} & \textbf{4.20\%} & \textbf{0.91} & \textbf{0.99} \\
\quad w/o CoT & 15.1\% & 1.04 & 0.87 \\
\quad w/o CoT (Transcript Only) & 5.10\% & 0.94 & 0.92 \\
\quad w/o Acoustic Priors & 8.5\% & 0.95 & 0.74 \\
\bottomrule
\end{tabular}
\vspace{-0.3cm}
\end{table}

% ---------------------------------------------------------------
% 9. Conclusion
% ---------------------------------------------------------------
\section{Conclusion}
\label{sec:conclusion}
We present CogSR, a generative framework that synergizes the stability of Flow Matching with the cognitive capabilities of Large Audio-Language Models. By utilizing structured Chain-of-Thought reasoning as a semantic anchor alongside explicit multi-factor acoustic priors, our approach effectively bridges the gap between high-level understanding and low-level signal reconstruction. This design proves critical for restoring speech in severely bandwidth-limited environments, where traditional methods typically fail. Extensive experiments confirm that CogSR establishes a new state-of-the-art in both objective metrics and subjective perceptual fidelity.
% ---------------------------------------------------------------
% References
% ---------------------------------------------------------------
\bibliographystyle{IEEEtran}
\bibliography{icme2026references}

\end{document}